\documentclass[preprintnumbers, sort&compress, floatfix]{revtex4}
\usepackage{amsmath}
\usepackage{amsfonts}
\usepackage{amssymb}
\usepackage{graphicx}
\usepackage{bm}
\usepackage{hyperref}
\def\be{\begin{equation}}
\def\ee{\end{equation}}
\def\bi{\bibitem}

\begin{document}
\title{Noether Symmetries of a Modified Model in Teleparallel Gravity and a New Approach for Exact Solutions.}
\author{‪Behzad Tajahmad$ $}
\email{behzadtajahmad@yahoo.com}
\affiliation{Faculty of Physics, University of Tabriz, Tabriz, Iran}
\begin{abstract}
In this paper, we have presented the Noether symmetries of flat FRW spacetime in the context of a new action in Teleparallel Gravity which we construct it based on $f(R)$ version. This modified action contains a coupling between scalar field potential and magnetism. Also, we introduce an innovative approach (B.N.S. Approach) for exact solutions which carry more conserved currents than Noether approach. By data analysis of the exact solutions, obtained from Noether approach, late time acceleration and phase crossing are realized, and some deep connections with observational data such as the age of the universe, the present amounts of the scale factor as well as the state and deceleration parameters are observed. In B.N.S. approach, we have considered dark energy dominated era.
\\ \\
\textbf{Keywords}: Noether symmetry, Teleparallel gravity, B.N.S. Approach
\end{abstract}
\maketitle
\section{\bf{Introduction}}
In the last decade, one of the big challenges for physicists is the explanation of the essence and mechanism of the acceleration of our universe \cite{1}-\cite{3}, in this era of the universe, which has confirmed by some observation data such as supernova type Ia \cite{4}-\cite{5} baryon acoustic oscillations \cite{6} weak lensing \cite{7} and large scale structure \cite{8}. However, a plausible elucidation for this is commonly done using the model of a very exotic fluid called dark energy which has negative pressure. Another well-known possibility is to modify Einstein's general relativity \cite{9}, making the action of the theory dependent on a function of the curvature scalar $R$, but at a certain limit of parameters, the theory falls in general relativity. This procedure of explaining the accelerated expansion of our universe is known as modified gravity. An alternative, consistently describing the gravitational interaction, is one which only acknowledges the torsion of spacetime, thus canceling out any effect of the curvature. This approach is known as Teleparallel theory \cite{10}-\cite{11} which is demonstrably equivalent to general relativity. Teleparallel gravity enables one to say that gravity is not due to curvature, but the torsion.

The choice of unknown functions, somewhat arbitrary, such as coupling functions and potentials in the equations of motion obtained from the point like lagrangian of the extended models has given rise to objections of fine tuning, the very problem whose solutions have been set out through inflationary theories. Therefore, it is desirable to have a path to derive the potential or at least some criteria for acceptable potentials. One such approach is based on Noether symmetry and was recently applied by S. Capozziello $et$ $al$. \cite{12}-\cite{15}, R. de Ritis $et$ $al$. \cite{16}-\cite{17}, A. K. Sanyal $et$ $al$. \cite{18}-\cite{25}, and others \cite{26}-\cite{42}. Noether approach, representing several conserved currents (Noether currents), is not conducive to any solutions while matching all or a portion of with field equations. The more currents there are the more problems pile up. On the other hand, hidden currents derivable from continuity equation \cite{43}-\cite{44}, are desired to be included, but when doing so things get worse due to the abundance of currents. The point is, with or without intervening hidden currents, one is compelled to cross out some in order to reach at a solution. In this paper, we introduce a new approach, B.N.S. approach, in section \ref{IV} which keeps the maximum possible number of conserved currents including Noether, hidden and arbitrary ones.\\

This paper is organized as follows. In section \ref{II} we introduce the model and extract the point like lagrangian. In section \ref{III} we gently present the Noether symmetries, invariants and exact solutions of our model. Moreover, by data analysis, we demonstrate that the observational data corroborate our findings. In section \ref{IV} we introduce B.N.S. approach and study our model with it, especially in dark energy dominated era. In section \ref{V} we consider corresponding WDW-equation and eventually in section \ref{VI} we conclude the results.
\section{The model \label{II}}
In some articles such as Refs. \cite{45}-\cite{48}, the gravitational action
\begin{equation*}
S = \int d^4x \sqrt{-g} \Bigg[{{M_{PL}^{2}}\over 2} R + {1\over 2}\phi_{,\mu}\phi^{,\mu} - V(\phi) - {1\over 4}{f(\phi)^2}F_{\mu
\nu}F^{\mu \nu} \Bigg],
\end{equation*}
was investigated the studies of which led to satisfactory results (inflation, late--time--accelerated expansion, ...). Indeed, this action is the most generic action for a single field inflation. Gauge fields are the main driving force for the inflationary background. It is worth to note that there are several fields such as the vector fields and the nonlinear electromagnetic fields which are able to produce the negative pressure effects. In some papers such as Refs. \cite{47}-\cite{48}, the authors used this model, perhaps, to answer the question that whether or not this model may describe the late-time-accelerated expansion. Maybe, the main motivations for applying such models are the efforts for reaching at a unified model (with a single scalar field) which describes the stages of cosmic evolution. Anyway, such discussions are beyond the scope of this paper. Now, the $T$-version (Teleparallel theory with $T$) of this action is considered completely. Hence, we have
\be \label{action}S = \int d^4x e \Bigg[{{M_{PL}^{2}}\over 2} T + {1\over 2}\phi_{,\mu}\phi^{,\mu} - V(\phi) - {1\over 4}{f(\phi)^2}F_{\mu
\nu}F^{\mu \nu} \Bigg]\ee
where $e=\det(e_{\nu}^{i})=\sqrt{-g}$ with $e_{\nu}^{i}$ being a vierbein (tetrad) basis, $T$ is the torsion scalar, $\phi_{,\mu}$ stands for the components of the gradient of $\phi$ and $V(\phi)$ is the scalar field potential. The vector potential $\mathbf{A}$ of electromagnetic theory generates the electromagnetic field tensor via the geometric equation $\mathbf{F} = - \text{(antisymmetric part of $\mathbf{\nabla A}$)} $. Hence, for a given 4-potential $ A_{\mu} $, the field strength of the vector field is defined by $F_{\mu \nu}= \partial_{\mu} A_{\nu} - \partial_{\nu} A_{\mu} \equiv A_{\nu, \mu} - A_{\mu , \nu}$. As seen, in the action (\ref{action}) the gauge kinetic function $f^2(\phi)$ is coupled to the strength tensor $F_{\mu \nu}$.

In the flat FRW line element,
\be \label{frw}ds^2 = dt^2 - a^2(t) \left(dx^2+dy^2+dz^2\right),\ee
the scalar torsion takes the form $T=-6\dot{a}^2/{a^2}$ \cite{51}-\cite{52} where $a$ is the scale factor of the universe which depends on time only, and the dot denotes a derivative with respect to time.

Regarding (\ref{frw}), if we introduce the homogeneous and isotropic vector field as
\be \label{four vector}
A_{\mu} = (A_{0}; A_{1}, A_{2}, A_{3}) =(\chi (t); \frac{A(t)}{\sqrt{3}}, \frac{A(t)}{\sqrt{3}}, \frac{A(t)}{\sqrt{3}}),
\ee
then
\be\label{tensor}
F_{\mu \nu}F^{\mu \nu}= \frac{- 2\dot{A}^2}{a^2}
\ee
would be the case. However, one can choose the gauge $A_{0} = \chi(t) = 0$, by using the gauge invariance \cite{45}. Our background (FRW) implies $A_{1} = A_{2} = A_{3}$, hence we have no worry about changing the direction of the vector field in time. The action (\ref{action}) can be written in the canonical form $i.e.$ $S=\int dt L(Q,\dot{Q})+\Sigma_{0}$,
\be\label{point like in cosmic gauge}
L(Q,\dot{Q})= -3a\dot{a}^2+{1\over2}a^3\dot{\phi}^2+{1\over2}a {f(\phi)^2}\dot{A}^2 - a^3 V(\phi),
\ee
\begin{equation*}
  \text{surface-term} = \Sigma_{0} = 0,
\end{equation*}
where configuration space is $Q=(a,\phi,A)$ with tangent space $TQ=(a,\phi,A, \dot{a}, \dot{\phi}, \dot{A})$. We set the reduced Planck mass, $M_{PL}$, equal to one.
\section{Noether Symmetry \label{III}}
In this section, we study Noether symmetry approach for the action (\ref{action}). We split this section into two subsections. In first subsection \ref{A}, we study the general form of Noether symmetry which is perceived as Noether gauge symmetry. However, this terminology is wrong because there is no gauge \cite{48}-\cite{50}. In second subsection \ref{B}, we study spatial Noether symmetry which is distinguished as common Noether Symmetry.
\subsection{Noether Symmetry (NS): A General Approach\label{A}}
The Euler-Lagrange equations for a dynamical system are
\begin{equation*}
\frac{\partial L}{\partial q_{i}} - \frac{d}{dt} \left(\frac{\partial L}{\partial \dot{q}_{i}} \right) =0,
\end{equation*}
where $q_{i}$ are the generalized positions in the corresponding configuration space (i.e. $Q=\{q_{i}\}$). The energy function associated with the Lagrangian is given by
\begin{equation*}
E_{L} = \sum_{i}\dot{q}_{i}\frac{\partial L}{\partial \dot{q}_{i}}-L.
\end{equation*}
According to the point like lagrangian (\ref{point like in cosmic gauge}), the corresponding Euler-Lagrange equation for the scale factor $a$ becomes
\be \label{FE1}
3 \dot{a}^2 + \frac{3 a^2 \dot{\phi}^2}{2} + \frac{f^2 \dot{A}^2}{2} - 3 a^2 V + 6 a \ddot{a} = 0.
\ee
For the scalar field $\phi$, the Euler-Lagrange equation takes the following form
\be \label{FE2}
f f^\prime \dot{A}^2 - a^2 V^\prime - 3 a \dot{\phi} \dot{a} - a^2 \ddot{\phi}=0,
\ee
which is the Klein-Gordon equation. The prime indicates the derivative with respect to $\phi$. For the vector potential $A$, the Euler-Lagrange equation reads
\be \label{FE3}
\dot{a} f^2 \dot{A} + 2 a f f^\prime \dot{A} \dot{\phi} + a f^2 \ddot{A}=0.
\ee
And finally, the Hamiltonian constraint or total energy $E_{L}$ corresponding to (0,0)-Einstein equation becomes
\be \label{FE4}
\frac{ a^2 \dot{\phi}^2}{2} + \frac{ f^2 \dot{A}^2}{2} + a^2 V - 3 \dot{a}^2 = 0.
\ee
The dynamics of our system is given by these four equations.\\

The pressure and energy density of the scalar field can be written as
\be \label{P and rho}\begin{split}
P_{\phi} = \frac{\dot{\phi}^2}{2} - V(\phi) + \frac{f^2(\phi) \dot{A}^2}{6 a^2},\\
\rho_{\phi} = \frac{\dot{\phi}^2}{2} + V(\phi) + \frac{f^2(\phi) \dot{A}^2}{2 a^2},\\
\end{split} \ee
Consequently, the EoS parameter for a scalar field could be
\be \label{EoS}
W_{eff} = \frac{P_{\phi}}{\rho_{\phi}} = \frac{\frac{\dot{\phi}^2}{2} - V(\phi) + \frac{f^2(\phi) \dot{A}^2}{6 a^2}}{\frac{\dot{\phi}^2}{2} + V(\phi) + \frac{f^2(\phi) \dot{A}^2}{2 a^2}}\\
\ee

To solve the field Eqs. (\ref{FE1}) - (\ref{FE3}) we use the Noether symmetry approach.\\
A vector field
\be \label{vector field}
\textbf{X} = \xi(t,a,\phi,A)\frac{\partial}{\partial t}+\alpha(t,a,\phi,A)\frac{\partial}{\partial a}
+\beta(t,a,\phi,A)\frac{\partial}{\partial\phi} +\gamma(t,a,\phi,A)\frac{\partial}{\partial A},\ee
is a Noether symmetry of the Lagrangian (\ref{point like in cosmic gauge}), if there exists a vector valued function, $ G(t, a, \phi,
A)\in \Gamma $, where $ \Gamma $ is the space of differential functions such that
\be \label{noether equation}
\textbf{X}^{[\textbf{1}]} L + L  D_{t} \xi(t, a, \phi, A) = D_{t} G(t, a, \phi, A)
\ee
in which $ D_{t} $ given by
\be \label{total operator}
D_{t} \equiv \frac{\partial}{\partial t} + \dot{a}\frac{\partial}{\partial a}
+\dot{\phi}\frac{\partial}{\partial\phi} +\dot{A}\frac{\partial}{\partial A},
\ee
is the total derivative operator and $\textbf{X}^{[\textbf{1}]}$, first-order prolongation is defined by
\be \label{first prolongation}
\textbf{X}^{[\textbf{1}]} = \textbf{X} + (D_{t} \alpha - \dot{a} D_{t} \xi) \frac{\partial}{\partial \dot{a}} + (D_{t} \beta - \dot{\phi}
D_{t} \xi) \frac{\partial}{\partial \dot{\phi}} + (D_{t} \gamma - \dot{A} D_{t} \xi) \frac{\partial}{\partial \dot{A}}.
\ee
If \textbf{X} is the Noether symmetry corresponding to the Lagrangian (\ref{point like in cosmic gauge}), then
\be \label{conserved current}
\textbf{I} = \xi L + ( \alpha - \dot{a} \xi ) \frac{\partial L}{\partial \dot{a}} + ( \beta - \dot{\phi} \xi ) \frac{\partial L}{\partial
\dot{\phi}} + ( \gamma - \dot{A} \xi ) \frac{\partial L}{\partial \dot{A}} - G ,
\ee
is a conserved quantity associated with \textbf{X}. The Noether symmetry condition for the lagrangian (\ref{point like in cosmic gauge}) yields the following system of linear partial differential equations
\be \label{NS equation 1}
\xi _{a} = \xi _{\phi} = \xi _{A} = 0,
\ee
\be \label{NS equation 2}
G = - 3 \alpha a^2 V - a^3 V \xi _{t} - \beta a^3 V',
\ee
\be \label{NS equation 3}
G_{a} = - 6 a \alpha _{t},\quad G_{b} = a^3 \beta _{t}, \quad  G_{A} = a f^2 \gamma _{t},
\ee
\be \label{NS equation 6}\begin{split}
3 \alpha + 2 a \beta_{\phi} - a \xi_{t} = 0,
\end{split}\ee
\be \label{NS equation 7}\begin{split}
\alpha f + 2 \beta a f' + 2 a f \gamma_{A} - a f \xi_{t},
\end{split}\ee
\be \label{NS equation 8}\begin{split}
- 2 \alpha_{a} + a \xi_{t} = 0, \quad  - 6 \alpha_{\phi} + a^2 \beta_{a} = 0,
\end{split}\ee
\be \label{NS equation 10}\begin{split}
f^2 \gamma_{a} - 6 \alpha_{A} = 0, \quad f^2 \gamma_{a} + a^2 \beta_{A} = 0,
\end{split}\ee
\subsubsection{\textbf{Zero G}}
Among the many sets of answers which we found, only $V(\phi)$ and $f(\phi)$ to be non-zero are noteworthy. Our solutions are as below
\be \label{NS sol set 4}\begin{split}
\beta = & \frac{-2 \sqrt{6}}{3} \left(\frac{c_{3} e^{- \sqrt{6}\phi / 4}}{a^{3/2}}-c_{1}\right),\quad \gamma = \frac{2 c_{1}}{3} A +
c_{5},\quad \alpha = \frac{-2}{3} \left(\frac{c_{3} e^{- \sqrt{6}\phi / 4}}{a^{1/2}}\right)+ \frac{c_{1}}{3} a, \\& \xi = c_{1}t
+c_{2}, \quad V(\phi) = V_{0} e^{- \sqrt{6}\phi / 2}, \quad f(\phi) = f_{0} e^{- \sqrt{6}\phi / 12}.
\end{split}\ee
Thus, symmetry generators, $X_{i}$, on tangent space turn out to be
\be \label{s g 4}\begin{split}
\textbf{X}_{1} = t \frac{\partial}{\partial t} \frac{2 \sqrt{6}}{3} + \frac{2 A}{3} \frac{\partial}{\partial A} + \frac{a}{3}
\frac{\partial}{\partial a} , \quad
\textbf{X}_{2} = \frac{\partial}{\partial t} , \quad
\textbf{X}_{3} = \frac{-2 \sqrt{6}}{3} \frac{e^{- \sqrt{6} \phi /4}}{a^{3/2}} \frac{\partial}{\partial \phi} - \frac{2}{3} \frac{e^{-
\sqrt{6} \phi /4}}{\sqrt{a}} \frac{\partial}{\partial a}. \quad
\textbf{X}_{4} = \frac{\partial}{\partial A}.
\end{split} \ee
So, the corresponding conserved currents are
\be \label{c c 4}\begin{split}
\textbf{I}_{1} = \frac{- t a^3 \dot{\phi}^2}{2} + \frac{t a f^2 \dot{A}^2}{2} - t a^3 V + 3 t a \dot{a}^2 - 2 a^2 \dot{a} + \frac{2
\sqrt{6} a^3 \dot{\phi}}{3} + \frac{2 a f^2 A \dot{A}}{3} \quad
\textbf{I}_{2}= 3 a \dot{a}^2 -& \frac{ a^3 \dot{\phi}^2}{2} + \frac{ a f^2 \dot{A}^2}{2} -  a^3 V,  \\
\textbf{I}_{3} = 4 e^{- \sqrt{6} \phi /4} \sqrt{a} \dot{a} - \frac{2 \sqrt{6}}{3} e^{- \sqrt{6} \phi /4} a^{3/2} \dot{\phi}, \quad
\textbf{I}_{4} = a f^2 \dot{A}.
\end{split}\ee

Now, we use cyclic variables associated with the Noether symmetry generator $\textbf{X}_{3}$ to simplify the system of equations. Note that $(\partial\textbf{I}_{4} /\partial t) \equiv$ Eq.(\ref{FE3}). The existence of Noether symmetry assures the presence of cyclic variables, say $(t, a, \phi, A) \rightarrow (s, w, u, v)$ such that the Lagrangian becomes cyclic in one of them ($w$ in our case). Cyclic variables can be found by defining a transformation $i: (t, a, \phi, A) \rightarrow (s, w, u, v)$ as an interior product such that $i_{\textbf{X}_{3}} ds = 0, i_{\textbf{X}_{3}} dw = 1, i_{\textbf{X}_{3}} du =
0$ and $i_{\textbf{X}_{3}} dv = 0$ or, put differently, we have the following equations
\be \label{cyclic-equations}\begin{split}
\xi \frac{\partial s(t, a, \phi, A)}{\partial t} + \alpha \frac{\partial s(t, a, \phi, A)}{\partial a} + \beta \frac{\partial s(t, a,
\phi, A)}{\partial \phi} + \gamma \frac{\partial s(t, a, \phi, A)}{\partial A} =0,\\
\xi \frac{\partial w(t, a, \phi, A)}{\partial t} + \alpha \frac{\partial w(t, a, \phi, A)}{\partial a} + \beta \frac{\partial w(t, a,
\phi, A)}{\partial \phi} + \gamma \frac{\partial w(t, a, \phi, A)}{\partial A} =1,\\
\xi \frac{\partial u(t, a, \phi, A)}{\partial t} + \alpha \frac{\partial u(t, a, \phi, A)}{\partial a} + \beta \frac{\partial u(t, a,
\phi, A)}{\partial \phi} + \gamma \frac{\partial u(t, a, \phi, A)}{\partial A} =0,\\
\xi \frac{\partial v(t, a, \phi, A)}{\partial t} + \alpha \frac{\partial v(t, a, \phi, A)}{\partial a} + \beta \frac{\partial v(t, a,
\phi, A)}{\partial \phi} + \gamma \frac{\partial v(t, a, \phi, A)}{\partial A} =0,\\
\end{split} \ee
where (according to our study $i.e.$ $ \textbf{X}_{3}$)
\be \label{cof. in cyc.}
\alpha = \frac{-2}{3} a^{-1/2} e^{- \sqrt{6} \phi /4} \quad , \quad \beta = \frac{-2 \sqrt{6}}{3} a^{-3/2} e^{- \sqrt{6} \phi /4} \quad ,
\quad \gamma = \xi = 0,
\ee
$i.e.$ $c_{1} = c_{2} = c_{5} = 0$. The coordinate transformation (\ref{cyclic-equations}) is not unique and a clever choice can be
very advantageous. Moreover, the solution of Eq. (\ref{cyclic-equations}) is, in general, not defined on the entire space but only locally. Among the many sets of solutions which we found, we choose these solutions
\be \label{sol. for cyc.}
s = t \quad , \quad w = \frac{1}{2} a^{3/2} e^{- \sqrt{6} \phi / 4} \quad , \quad u = \frac{1}{2} a^{3/2} e^{ \sqrt{6} \phi / 4} \quad ,
\quad v = A.
\ee
Therefore we have $(t, a, \phi, A) \rightarrow \left(t, \frac{1}{2} a^{3/2} e^{- \sqrt{6} \phi / 4}, \frac{1}{2} a^{3/2} e^{ \sqrt{6}
\phi / 4}, A \right)$ in which $w$ is a cyclic variable. Thus, the scale factor, scalar field, coupling function, and the scalar field
potential can be written as
\be \label{inv. sol. for cyc.}
a = (4 u w)^{1/3} \quad , \quad \phi = \frac{2}{\sqrt{6}} \ln\left(\frac{w}{u}\right) \quad , \quad f(\phi) = f_{0} \left(\frac{u}{w}
\right)^{1/6} \quad , \quad V(\phi) = \frac{V_{0} u}{w}.
\ee
The point-like lagrangian (\ref{point like in cosmic gauge}) in terms of new variables, then reads
\be \label{lag. for cyc.}
\mathcal{L} = \mathcal{L}(u, w, A, \dot{u}, \dot{w}, \dot{A})= \frac{-16}{3} \dot{u} \dot{w} - 4 V_{0} u^2 + 2^{-1/3} f_{0}^2 u^{2/3}
\dot{A}^2.
\ee
The Euler-Lagrange equations lead to
\be \label{Eu.lag.Eq. for cyc.}
\ddot{u} = 0 \quad , \quad 2^{2/3} f_{0}^2 \dot{A}^2 - 24 V_{0} u^{4/3} + 16 \ddot{w} u^{1/3} = 0 \quad , \quad 3 u \ddot{A} + 2 \dot{A}
\dot{u} = 0,
\ee
and the corresponding conserved current ($\textbf{I}_{3}$) will be
\be \label{Eu.lag.Eq. for cyc. I3}
\textbf{\~{I}}_{3} = \frac{16}{3} \dot{u}\qquad \rightarrow \qquad \ddot{u} = 0.
\ee
As we observe, this equation doesn't add any new equation. Solutions for the Eq.(\ref{Eu.lag.Eq. for cyc.}) are
\be \label{Sol. for.Eu.lag.Eq. for cyc.}\begin{split}
u(t) &= c_{5} t + c_{6},\\
A(t) &= c_{3} + c_{4} \int {\frac{dt}{u^{2/3}}} = c_{3} + \frac{3 c_{4} (c_{5}t+c_{6})^{1/3}}{c_{5}} + c_{7},\\
w(t) &= \int \left[ \int \left(\frac{24 V_{0} u^{4/3} - 2^{2/3} f_{0}^2 \dot{A}^2}{16 u^{1/3}} \right)dt \right]dt + c_{1} t +c_{2}\\
&= \frac{1}{32 c_{5}^2} \left[ (2^{1/3} 3 c_{4} f_{0})^2 (c_{5}t + c_{6})^{1/3} + 8 V_{0} (c_{5} t + c_{6})^3 \right] + (c_{8} + c_{1}) t
+ c_{2} + c_{9},
\end{split} \ee
where $\{ c_{i}$ $;$ $ i = 1 , ... ,9 \}$ are constants of integration. Inserting values of $u(t)$, $A(t)$, and $w(t)$ in Eq.(\ref{inv.
sol. for cyc.}), we obtain

\be \label{inv. inv. sol. for cyc.1.}\begin{split}
a(t) &= \left[ \left(\frac{ 3 f_{0} c_{4}}{2^{7/6} c_{5}}\right)^2 (c_{5}t + c_{6})^{4/3} + \frac{V_{0}}{c_{5}^2} (c_{5}t + c_{6})^4 + 4
(c_{5} t + c_{6})\{(c_{1}+c_{8}) t + c_{2} + c_{9} \} \right]^{1/3},
\end{split} \ee
\be \label{inv. inv. sol. for cyc.2.}\begin{split}
\phi(t) &= \frac{\sqrt{6}}{3} \ln \left[ \left(\frac{3 f_{0} c_{4}}{2^{13/6} c_{5}} \right)^2 (c_{5} t + c_{6})^{-2/3} + \frac{V_{0}}{4
c_{5}^2} (c_{5}t + c_{6})^2 + (c_{5} t + c_{6})^{-1} \{(c_{1} + c_{8})t + c_{2} + c_{9}\} \right],
\end{split} \ee
\be \label{inv. inv. sol. for cyc.3.}\begin{split}
f(\phi)&= f_{0}\exp\left[ \frac{\sqrt{6} \phi}{-12} \right] = f_{0} N(\phi) \\ &\equiv f_{0} \left[ \frac{32 c_{5}^2 (c_{5} t + c_{6} )}{ (2^{1/3} 3 c_{4} f_{0})^2 (c_{5}t + c_{6})^{1/3} + 8 V_{0} (c_{5} t +
c_{6})^3 + 32 c_{5}^2(c_{8} + c_{1}) t + 32 c_{5}^2 ( c_{2} + c_{9}) } \right]^{1/6} = f(t),
\end{split} \ee
\be \label{inv. inv. sol. for cyc.4.}\begin{split}
V(\phi) &= V_{0} \exp \left[\frac{\sqrt{6} \phi}{-2} \right] \\ &\equiv V_{0}\left[ \frac{32 c_{5}^2 (c_{5} t + c_{6} )}{ (2^{1/3} 3 c_{4} f_{0})^2 (c_{5}t + c_{6})^{1/3} + 8 V_{0} (c_{5} t + c_{6})^3
+ 32 c_{5}^2(c_{8} + c_{1}) t + 32 c_{5}^2 ( c_{2} + c_{9}) } \right] = V(t).\\ \\
\end{split} \ee

To see the behavior of important quantities with these solutions, we choose constants as below
\be \label{selection for constants}\begin{split}
c_{1} = 0.2795084975 \qquad c_{2} = 1.615579240 \times 10^{10} \qquad c_{4} = 1.120082910 \times 10^{-14} \qquad \\
c_{5} = 1.118033986 \times 10^{-21} \qquad f_{0} = \sqrt{-1} = i \qquad V_{0} = 10.00000003 \qquad
c_{3} = c_{6} = c_{7} = c_{8} = c_{9} = 0.
\end{split} \ee
Perhaps, the\label{222} amount of $f_{0}$ seems strange, but the square of $f_{0}$ matters in the action (\ref{action}) and also in the relevant equations, so it is not problematic.
\begin{figure}
\centering
\includegraphics[width=17 cm, height=2.2 in]{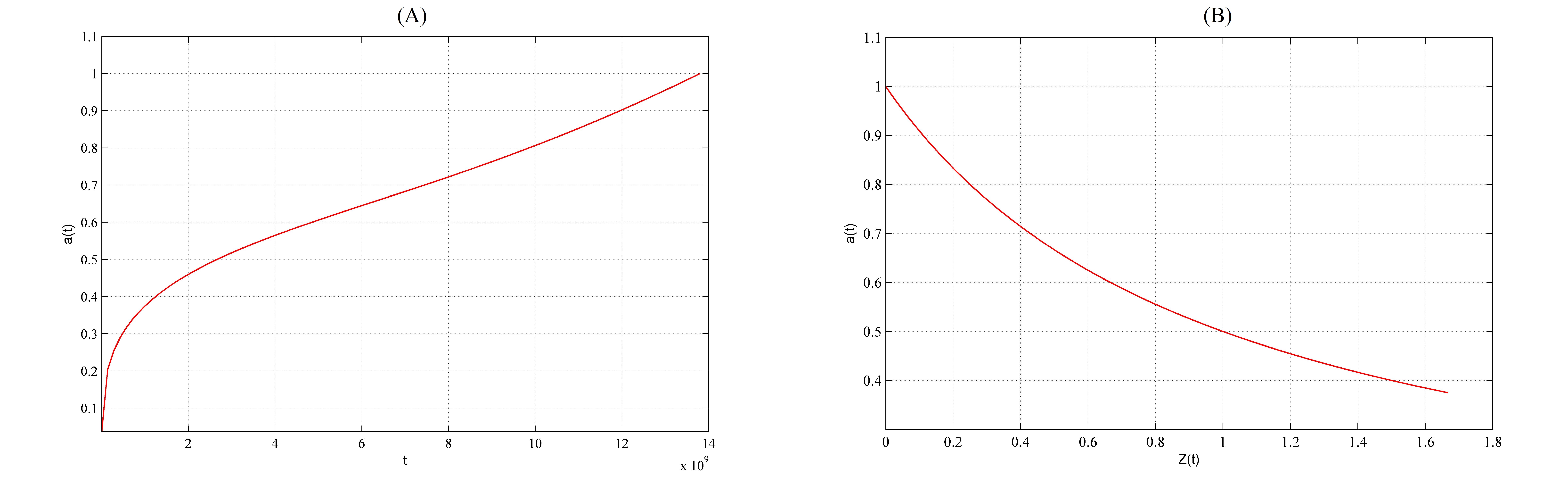}\\
\caption{Plot (A) indicates the scale factor $a(t)$ versus time $t$ at the time range [0.7Myr , 13.8Gyr] while (B) shows the scale factor $a(t)$ versus redshift $z$ at the time range [1Gyr , 13.8Gyr]}\label{aa}
\end{figure}
\begin{figure}
\centering
\includegraphics[width=17 cm,  height=2.2 in]{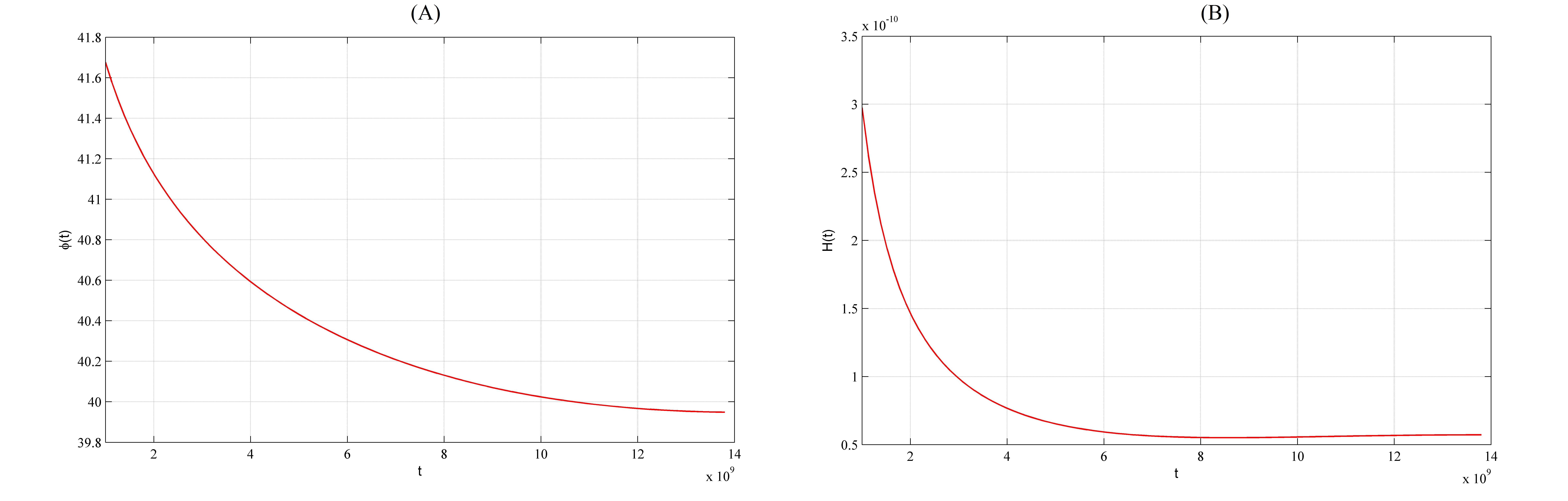}\\
\caption{Plots (A) and (B) indicate the scalar field $\phi(t)$ and Hubble parameter $H(t)$ versus time $t$ at the time range [1Gyr , 13.8Gyr], respectively.}\label{phiH}
\end{figure}
\begin{figure}
\centering
\includegraphics[width=17 cm, height=2.2 in]{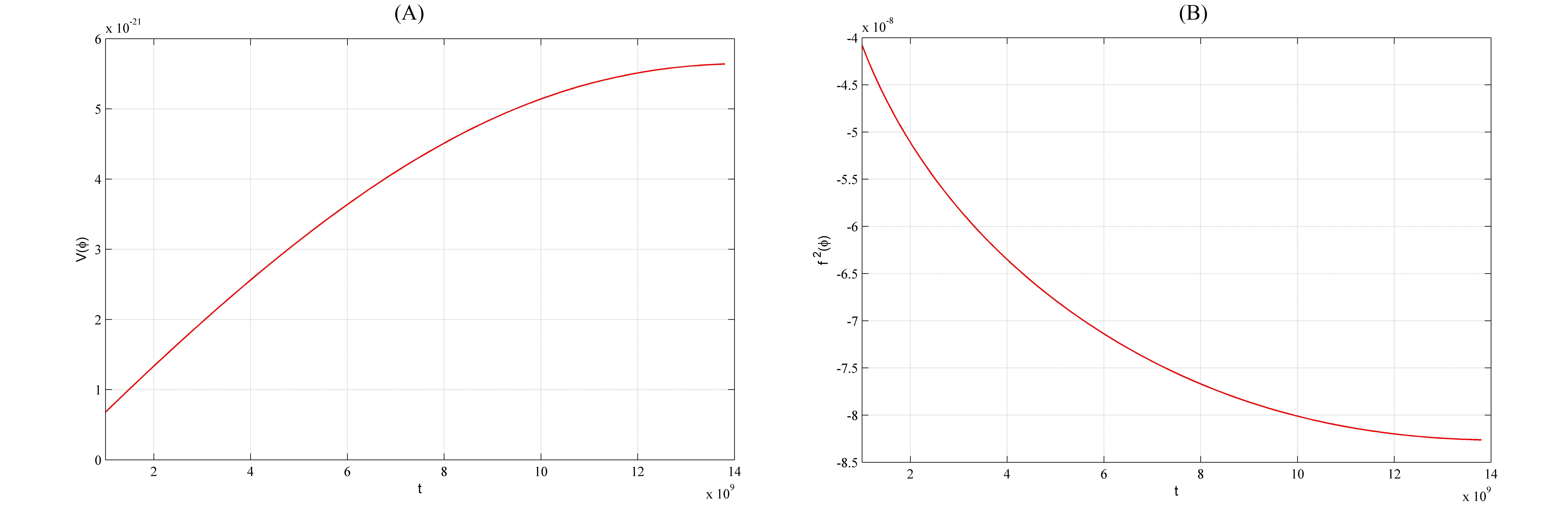}\\
\caption{Plots (A) and (B) indicate the scalar potential $V(\phi)$ and coupling function $f^{2}(\phi)$ versus time $t$ at the time range [1Gyr , 13.8Gyr] respectively.}\label{Vf2}
\end{figure}
\begin{figure}
\centering
\includegraphics[width=17 cm, height=2.2 in]{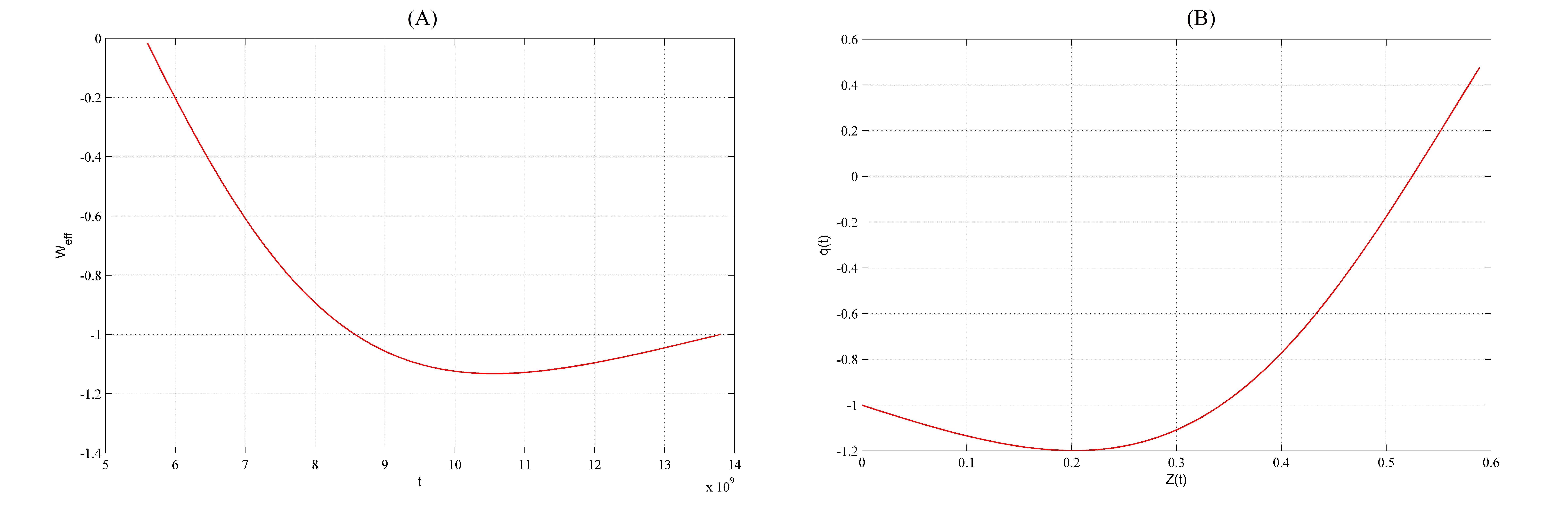}\\
\caption{Plots (A) and (B) indicate the state parameter $W_{eff}$ and deceleration parameter $q$ versus redshift $z$ at the time range [5.6Gyr , 13.8Gyr] respectively.}\label{wq}
\end{figure}
We present four figures with data analysis. Figure \ref{aa}(A) indicates the scale factor with increasing nature expressing first the decelerated and then the accelerated expansion of the universe. The present values of the age of the universe and the scale factor are $t_{0} = 13.80$ $Gyr$ and $ a_{0} = 1.00$, respectively. Figure \ref{aa}(B) indicates that the redshift goes down while the scale factor increases with time. Present value of the redshift is $z_{0}= 0$. Figures \ref{phiH}(A) and \ref{phiH}(B) show the scalar field and Hubble parameter with decreasing natures versus time, as we expect. The present values of the Hubble parameter and scalar field are $H_{0} = 5.7212 \times 10^{-11} yr^{-1} \equiv 55.98 km.s^{-1}.Mps^{-1}$ and $\phi_{0} = 39.95$. We do not present plots of the scalar potential $V(\phi)$ and coupling function $f^{2}(\phi)$ versus $\phi$ because their behaviors are obvious (decreasing exponentially). In Figure (\ref{Vf2}) $V(\phi)$ and $f^2(\phi)$ have been plotted with respect to time. Figure \ref{Vf2}(A) shows the scalar potential increases with time while Figure \ref{Vf2}(B) indicates the coupling function $f^{2}(\phi)$ decreasing with time, however, the absolute value of $f^2(\phi)$ is increasing such as $V(\phi)$ vs time. Astrophysical data show that $W_{eff}$ lies in a very narrow band close to $W_{eff} = -1$. The behavior of $W_{eff}$ in Figure \ref{wq}(A) indicates that the crossing of the phantom divide line $W_{eff} = -1$ occurs from the quintessence phase $W_{eff} > -1$ to the phantom phase $W_{eff} < -1$. The present value of the EoS parameter is calculated to be $ W_{eff_{0}} = -1.00$. The deceleration parameter, $q = -(a \ddot{a}) / \dot{a}^2$, shows first positive, indicative of decelerating universe, and then negative behavior, implying an accelerating universe (See Figure. \ref{wq} (B)). So, late--time--accelerated expansion is realized. The present value of the deceleration parameter is measured to be $q_{0} = -1.00$, and at the time $t_{ac.}= 6.294 Gyr$, we have $q(t_{ac.}) = 0$, so, acceleration starts at the redshift value, $z(t_{ac.}) = 0.524$, which is at about half the age of the universe.\\

\begin{center}
\item \textbf{$\bullet$ Satisfaction of Maxwell's Equations.}
\end{center}

Here, we want to answer the question that with the obtained form of the vector potential, are Maxwell's equations satisfied or not. For this purpose, we must utilize Maxwell's equations in curved space-time which in terms of components of the field tensor $\mathbf{F}$ are as \cite{54}
\begin{equation}\label{M1}
F_{\alpha \beta , \gamma}+F_{\beta \gamma , \alpha} + F_{\gamma \alpha , \beta} =0,
\end{equation}
\begin{equation}\begin{split}\label{M2}
{F^{\alpha \beta}}_{,\beta} = -4 \pi J^{\alpha}\qquad; \qquad
\left\{
\begin{array}{ll}
\text{if $\alpha = 0$}:& \hbox{$J^0 = \rho = \text{charge density,}$} \\ \\
\text{if $\alpha \neq 0$}:& \hbox{$(J^1, J^2, J^3) = \text{components of current density,} $}
\end{array}
\right. \end{split}
\end{equation}
where $\{J^{\alpha}$ ; $\alpha \in \{0, 1, 2, 3\}\}$ are the components of the 4-current $\mathbf{J}$. In a nutshell, through Eq. (\ref{M1}) magnetodynamics and magnetostatics, and through Eq. (\ref{M2}) electrodynamics and electrostatics are unified in one geometric law. The usual form of Maxwell's equations may be reached at easily since Eq. (\ref{M1}) reduces to $\mathbf{\nabla} \mathbf{\cdot} \mathbf{B} = 0$ when one takes $\alpha = 1$, $\beta = 2$, $\gamma = 3$; and it reduces to $\partial \mathbf{B} / \partial t + \mathbf{\nabla} \times \mathbf{E} = 0$ when one sets any index, e.g., $\alpha = 0$, and finally, with Eq. (\ref{M2}) two of Maxwell's equations, $\mathbf{\nabla \cdot E }= 4 \pi \rho$ (the electrostatic equation), $\partial \mathbf{E} / \partial t - \mathbf{\nabla} \times \mathbf{B} = -4 \pi \mathbf{J}$ (the electrodynamic equation), are obtained by putting $\alpha = 0$ and $\alpha \neq 0$, respectively. \\

For the electromagnetism part of the action (\ref{action}), i.e.
\begin{equation}\label{M3}
\mathcal{L}_{EM} = -\frac{1}{4}\int d^4 x \sqrt{-g}f(\phi)^2 F_{\mu \nu}F^{\mu \nu} = -\frac{1}{4}\int d^4 x \sqrt{-g} g^{\alpha \beta} g^{\mu \nu} f(\phi)^2 F_{\mu \alpha} F_{\nu \beta},
\end{equation}
Eqs. (\ref{M1}) and (\ref{M2}) read
\begin{equation}\label{M41}
\partial^{\alpha} \left(\sqrt{-g} f^2 F^{\beta \gamma} \right) + \partial^{\beta} \left(\sqrt{-g} f^2 F^{\gamma \alpha} \right) +\partial^{\gamma} \left(\sqrt{-g} f^2 F^{\alpha \beta} \right) =0,
\end{equation}
\begin{equation}\label{M42}
\partial _{\mu} \left[\sqrt{-g} g^{\alpha \beta} g^{\mu \nu} f^2(\phi) F_{\nu \beta}\right] =0 \qquad \longrightarrow \qquad \left(\sqrt{-g}f^2(\phi) F^{\alpha \mu} \right)_{,\mu} = 0,
\end{equation}
respectively. Note that in our studying case, we have $\mathbf{J}=(J^0, J^1, J^2, J^3) = (0, 0, 0, 0)$. After simplifying, both equations (\ref{M41}, \ref{M42}) lead to the same equation, viz,
\begin{equation}\label{Mo}
\frac{\partial}{\partial t} \left(a f^2 \dot{A} \right) = 0.
\end{equation}
Clearly, this equation is equivalent to the third field equation (i.e. Eq.(\ref{FE3})). Hence, Eqs. (\ref{M1}) and (\ref{M2}) have been satisfied automatically when the solution for the field equations was found. Therefore, the results are consistent with all Maxwell's field equations.\\

Now, let us define the electric $\mathbf{E}$ and magnetic $\mathbf{B}$ fields covariantly which are seen by an observer who is characterized by the 4-velocity vector $u^{\mu}$. One has \cite{55}
\begin{equation}\label{eb1}
E_{\mu} = u^{\nu} F_{\mu \nu} , \qquad B_{\mu} = \frac{1}{2} \varepsilon_{\mu \nu \kappa} F^{\nu \kappa},
\end{equation}
where the tensor $\varepsilon_{\mu \nu \kappa}$ is defined by the relation
\begin{equation}\label{eb2}
\varepsilon_{\mu \nu \kappa} = \eta_{\mu \nu \kappa \lambda} u^{\lambda},
\end{equation}
in which $\eta_{\mu \nu \kappa \lambda}$ is an antisymmetric permutation tensor of space-time with $\eta^{0123} = 1/\sqrt{-g}$ or $\eta_{0123} = \sqrt{-g}$. In cosmic time for a comoving observer with $u^{\mu}= (1, 0, 0, 0)$, we get
\begin{equation}\begin{split}\label{eb3}
E_{\mu} =
\left\{
\begin{array}{ll}
\text{$E_{i} = - \dot{A}_{i}$} \quad;\quad& \hbox{for $\mu = i =1, 2, 3$} \\ \\
\text{$0$}\quad;\quad& \hbox{for $\mu = 0$}
\end{array}
\right.\qquad , \qquad
B_{\mu} =
\left\{
\begin{array}{ll}
\text{$B_{i} = \frac{1}{a} \epsilon_{ijk} \partial_{j} A_{k}$} \quad;\quad& \hbox{for $\mu =i, j, k \in \{1, 2, 3\}$} \\ \\
\text{$0$}\quad;\quad& \hbox{Otherwise}
\end{array}
\right.\end{split}
\end{equation}
where $\epsilon_{ijk}$ is the well-known Levi-Civita symbol with $\epsilon_{123}=1$. In our case, we have obtained the form $A_{\mu} = (\chi(t), \theta t^{1/3}, \theta t^{1/3}, \theta t^{1/3})$ with $\theta = \sqrt{3} c_{4} c_{5}^{-2/3}$ for the 4-vector potential, hence we get
\begin{equation}\label{eb4}
E_{\mu} = \frac{\theta}{3} t^{-2/3} (0, 1, 1, 1) , \qquad B_{\mu} = \frac{\theta}{3 a} t^{-2/3} (0, 1, 1, 1).
\end{equation}
According to these obtained forms, both $E$ and $B$ decay with time. Data analysis show that the present amounts of these are $E_{\mu 0}= 3.13 \times 10^{-7}(0, 1, 1, 1)$ and $B_{\mu 0}= 3.13 \times 10^{-7}(0, 1, 1, 1)$, so their norm at present time is equal, i.e., $\| \mathbf{E }\|_{0} = \| \mathbf{B} \|_{0}$.
\subsubsection{\textbf{Non-Zero G}}
Because in our study, we focus on the non-constant form of $V(\phi)$ and $f(\phi)$, so these would lead to strange results for any choice of function $G(t, a, \phi, A)$ except zero. Therefore, by choosing non-zero $G$-function, we will not have any conserved current, because all the Noether coefficients are zero.
\subsection{Spatial Noether Symmetry (SNS)\label{B}}
Obviously, for getting the SNS-equations, we must take $ G = \xi = 0 $. Our solutions for these equations are as below
\be \label{SNS sol set 3}\begin{split}
\beta = \frac{ c_{1} c_{4} \sqrt{6}\left(c_{2} e^{\sqrt{6}\phi /4} - c_{3} e^{- \sqrt{6}\phi /4}\right)}{a^{3/2}}, \quad \gamma
=c_{8},\quad  \alpha = \frac{c_{1} c_{4} \left(c_{2} e^{\sqrt{6}\phi /4} + c_{3} e^{- \sqrt{6}\phi /4}\right)}{a^{1/2}},\\ f(\phi) =
c_{9} e^{\sqrt{6}\phi /12} \left(c_{2} - c_{3} e^{- \sqrt{6} \phi /2} \right)^{1/3}, \quad V(\phi) = c_{7} \left(c_{3}^{2}
e^{-\sqrt{6} \phi /2} - 2 c_{2}c_{3} + c_{2}^{2} e^{\sqrt{6}\phi /2} \right).
\end{split}\ee
Symmetry generators, $X_{i}$, on tangent space, become
\be \label{s g s3}\begin{split}
\textbf{X}_{1} = \sqrt{\frac{6}{a}} e^{\sqrt{6} \phi /4} \left(\frac{\partial}{\partial a} + \frac{1}{a} \frac{\partial}{\partial
\phi}\right),\qquad
\textbf{X}_{2} = \sqrt{\frac{6}{a}} e^{- \sqrt{6} \phi /4} \left(\frac{\partial}{\partial a} - \frac{1}{a} \frac{\partial}{\partial
\phi}\right),\qquad
\textbf{X}_{3} = \frac{\partial}{\partial A},
\end{split} \ee
Consequently, the corresponding conserved currents take the forms
\be \label{c c s3}
\textbf{I}_{1} = \sqrt{6 a} e^{\sqrt{6} \phi /4} \left(-\sqrt{6} \dot{a} + a \dot{\phi} \right), \qquad
\textbf{I}_{2} = \sqrt{6 a} e^{- \sqrt{6} \phi /4} \left(-\sqrt{6} \dot{a} - a \dot{\phi} \right), \qquad
\textbf{I}_{3} = a f^2 \dot{A}.
\ee
By a little delicacy, one can find that this case is similar to NS case ($i.e$. the results will be the same). So, it's futile to follow this set of solutions. It's sufficient to say that these generators (and also conserved currents) correspond to $\textbf{X}_{3}$ and $\textbf{X}_{4}$ (so $\textbf{I}_{3}$ and $\textbf{I}_{4}$) in NS approach (See Eqs. (\ref{s g 4}) and (\ref{c c 4})).
\section{Beyond Noether symmetry Approach (B.N.S. Approach) \label{IV}}
In this section, we have innovated an approach for exact solutions. We named it to be ``B.N.S. Approach" (B.N.S. is an abbreviation for \textbf{B}eyond \textbf{N}oether \textbf{S}ymmetry). This approach is so useful for extended gravity because we have some degrees of
freedom. Also, we may have more conserved currents.
Let us explain this approach.\\
In any action of extended gravity, we have some unknown functions such as $V(\phi)$ and $f(\phi)$ in our case. As a regular way, we may use Noether approach for defining them. As we observe in our case, and also in almost all other cases in articles, we can't reach the solutions which carry \textbf{all} conserved currents or at least more of those. For solving the field equations, we have to remove some of the conserved currents. On the other hand, symmetries have always played a central role in the conceptual discussion of the classical and quantum physics. I found that the main problem is the form of these unknown functions, the main culprit in removing some of the conserved currents. In the case in which we have new forms of these unknown functions, then the problem can be solved. B.N.S. approach carries out it through a simple way. Suppose that $F_{1} (\varphi)$, $F_{2} (\varphi)$, ..., $F_{n} (\varphi)$  are unknown functions where $\varphi = \varphi(t)$. First of all, we list all field equations and possible conserved currents, then do the maps as follows:\\
\begin{center}\label{B.T.Map}
\begin{tabular}{c c c c }
1. & $F_{1}(\varphi) \rightarrow F_{1}(t)$, & $ $ $ $ $ $ $\text{So we have:}$ $ $ $ $ & $F_{1}^{\prime}(\varphi) \rightarrow \frac{\dot{F_{1}}(t)}{\dot{\varphi}(t)}, $\\ \\
2. & $F_{2}(\varphi) \rightarrow F_{2}(t)$, & $ $ $ $ $ $ $ $ $ $ & $F_{2}^{\prime}(\varphi) \rightarrow \frac{\dot{F_{2}}(t)}{\dot{\varphi}(t)},$ \\ \\
$\vdots$ & $ $ $ $&$\Longrightarrow$  & $\vdots$ \\ \\
$n$. & $F_{n}(\varphi) \rightarrow F_{n}(t)$, & $ $ $ $ $ $ $ $ $ $ & $F_{n}^{\prime}(\varphi) \rightarrow \frac{\dot{F_{n}}(t)}{\dot{\varphi}(t)} $ $ ,$ \\ \\
\end{tabular}
\end{center}
where the prime indicates a derivative with respect to $\varphi$, and the dot indicates differentiation with respect to time. By substituting these in all equations, we may solve our ODE-system easily. After solving the system, we do an inverse map for reaching at the usual form of unknown functions ($i.e.$ depending on $\varphi$). Perhaps, in some cases, the inverse map be hard to reach at. In such cases, one can do it numerically. In numerical inverse mapping, two options are in order demanding initial values or the time interval, only. Note that one can first carry out Noether approach for getting the conserved currents, and then proceed with this approach $i.e.$ D.E-system = \{Field equations + Noether Conserved Currents + Other conserved currents\} without paying any attention to the form of unknown functions which are obtained by Noether approach. So, one could see that the form of unknown function may be different from those which is derived from Noether approach. Finding the solutions which carry all founded conserved currents with Noether approach is challenging but with B.N.S.--approach this road is paved.

Now, we carry out this approach in our case.\\
For solving the field equations (Eq.\ref{FE1}, \ref{FE2} and \ref{FE3}), without any loss of generality, we do the maps as follows
\be\label{map 1-1} \begin{split}\left\{
\begin{array}{ll}
i:& \hbox{$f(\phi(t)) \rightarrow f(t),$} \\ \\
ii:& \hbox{$V(\phi(t)) \rightarrow V (t).$}
\end{array}
\right. \end{split} \ee
Therefore, we have
\be \label{map 1-2} \begin{split}\left\{
\begin{array}{ll}
i:& \hbox{$f^{\prime}(\phi)\rightarrow \frac{\dot{f}(t)}{\dot{\phi}(t)},$} \\ \\
ii:& \hbox{$V^{\prime}(\phi)\rightarrow \frac{\dot{V}(t)}{\dot{\phi}(t)}.$}
\end{array}
\right. \end{split} \ee
Let us add some other conserved currents from NS (Eq.\ref{c c 4}). We would like to add $\textbf{I}_{2}=0$, $\textbf{I}_{3} = 0$, and $\textbf{I}_{4} = c_{0}^2$, in which $c_{0}$ is a constant. Note that Eq. (\ref{FE4}) (Hamiltonian constraint) is $\textbf{I}_{2}$.
Substituting Eqs. (\ref{map 1-1}) and (\ref{map 1-2}) in Eqs. (\ref{FE2}), (\ref{FE3}) and $\partial \textbf{I}_{4} / \partial t = 0$, we obtain the modified system of differential equations\\ \\
\textbf{ODE-Sys.} = \textbf{\{}Eq. (\ref{FE1}) $\cup$ Eq. (\ref{FE2}) $\cup$ Eq. (\ref{FE3}) $\cup$ $\textbf{I}_{2}=0$ $\cup$ $\textbf{I}_{3} = 0$
$\cup$ $d\textbf{I}_{4} / dt = 0$ \textbf{\}}.\\ \\
Solving it leads to (Set - 1)
\be \label{sol-FE-b.t.1}\begin{split}
\phi(t)&= \text{Any arbitrary function of time.} = b , \\
a(t) &= c_{3} \exp\left(\frac{b}{\sqrt{6}}\right),\\
A(t)&= c_{1} + c_{2} \left( \int \left\{\exp \left[\int \left( \frac{3\sqrt{6} \dot{b}^3 + 2\sqrt{6} \dddot{b}+18 \dot{b} \ddot{b}}{6 \dot{b}^2 + 2\sqrt{6} \ddot{b}} \right)dt \right] \right\}dt \right) ,\\
f(t) &= \frac{a \left(-2\left(3\dot{b}^2 + \sqrt{6} \ddot{b}  \right) \right)^{1/2}}{2 \dot{A}},\\
V(t) &= \frac{1}{4} \left( \ddot{b} + 3 \dot{b}^2 \right).\\
\end{split} \ee
If we do calculations with $\textbf{I}_{4} = c_{0}^2$, instead of $\partial \textbf{I}_{4} / \partial t= 0$ the results will be the same. limpidly, we have three symmetry generators ($\textbf{X}_{2}$, $\textbf{X}_{3}$ and $\textbf{X}_{4}$) with these solutions for the time being. We know that $\phi(t)$ must decay with time. According to Eq. (\ref{sol-FE-b.t.1}), if we set a function with decaying nature for $\phi(t)$, it leads to the same behavior for $a(t)$ as well. So, it is not admissible, hence we desert this solution. More conserved currents are the reason for this non-physical solution.

Removing one of the conserved currents, $\textbf{I}_{3} = 0$, the results are as below (Set - 2)
\be \label{sol-FE-b.t.2}\begin{split}
a(t)&= \text{Any arbitrary function of time.} = F_{1} \quad, \quad A(t)= \text{Any arbitrary function of time.}= F_{2},\\
f(t) &= c_{0} \left( F_{1} \dot{F_{2}} \right)^{-1/2},\\
\phi(t)&= c_{1} + \int \left[\frac{\sqrt{- 6 f^2 \dot{F_{2}^2} - 18 F_{1} \ddot{F_{1}} + 18 \dot{F_{1}^2}}}{ 3 F_{1}} \right] dt ,\\
V(t) &= \frac{-f^2 \dot{F_{2}}^2 + 6 F_{1} \ddot{F_{1}} + 12 \dot{F_{1}}^2}{6 F_{1}^2}
\end{split} \ee \\
where $f = f(t)$. As we observe, we have freedom degrees for choosing the form of $a(t)$ in both cases.\\
\begin{itemize}
  \item \textbf{An example for better understanding the analytical inverse map process.}
\end{itemize}
 In ``\textbf{Set - 1}", one can assume the following non-physical form for $b$
\begin{equation*}
b = \sqrt{6} t^2 \quad \rightarrow  \quad t^2 = \frac{\phi(t)}{\sqrt{6}}.
\end{equation*}
So, we obtain
\begin{equation*}
a(t) = \exp(t^2), \quad \quad A(t) = t \exp(3 t^2), \quad
\end{equation*}
\begin{equation*}
f(t)= \frac{\sqrt{-6 \left(6 t^2 + 1 \right)} \exp(-2 t^2)}{6 t^2 + 1} \quad \rightarrow \quad f(\phi) = \frac{\sqrt{-6 \left(\sqrt{6} \phi + 1 \right)} \exp\left(\frac{-2 \phi}{\sqrt{6}} \right)}{\sqrt{6} \phi + 1}\quad \rightarrow \quad f(\phi)^2 = \frac{-6 \exp \left( \frac{-2 \sqrt{6} \phi}{3}\right)}{\sqrt{6}\phi + 1},
\end{equation*}
\begin{equation*}
V(t) = 18 t^2 + 1 \quad \rightarrow \quad V(\phi) = \frac{18}{\sqrt{6}} \phi + 3.
\end{equation*}
\begin{figure}
\centering
\includegraphics[width=13 cm, height=2.4 in]{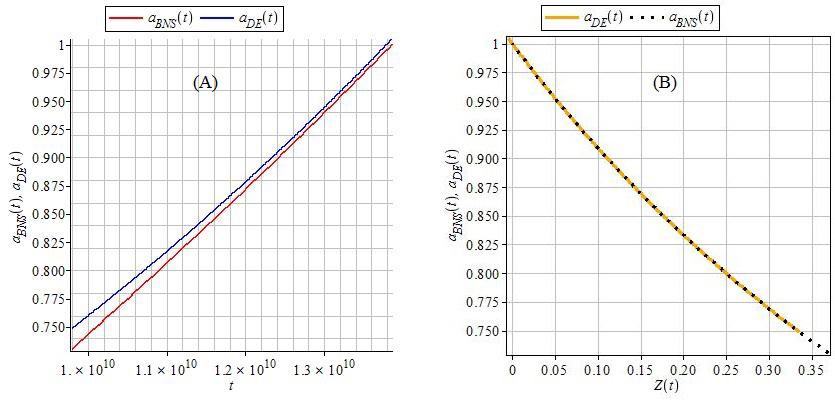}\\
\caption{Plot (A) indicates the scale factors $a_{D.E.}(t)$ and $a_{B.N.S.}(t)$ versus time $t$ at the time range [9.8Gyr , 13.8Gyr] while (B) shows both scale factors versus their own redshifts $z$ (i.e. $z_{D.E.}$ and $z_{B.N.S.}$) at the same time range evolving.}\label{fig5}
\end{figure}
\begin{figure}
\centering
\includegraphics[width=17 cm, height=2.4 in]{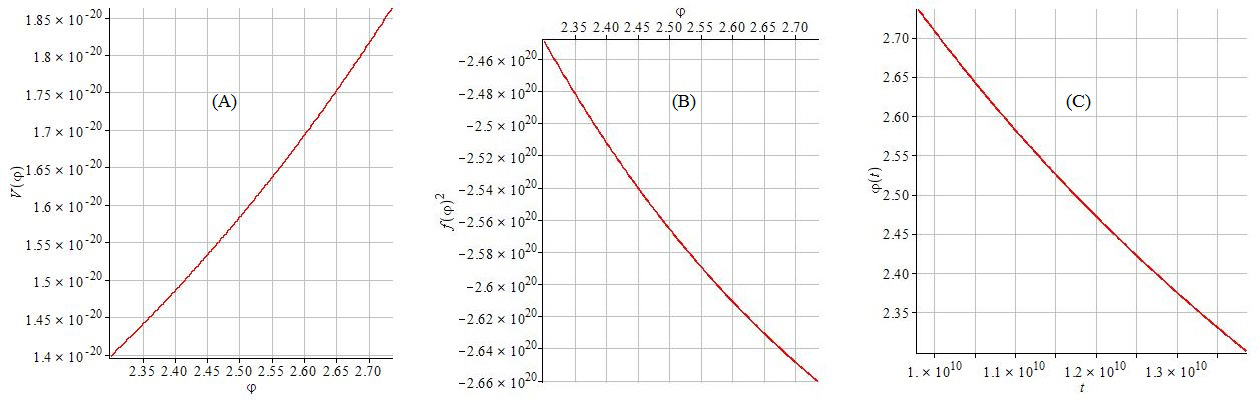}\\
\caption{Plots (A) and (B) indicate the scalar potential $V(\phi)$ and coupling function $f^2(\phi)$ versus scalar field $\phi$ at the time evolving range [9.8Gyr , 13.8Gyr] respectively while plot (C) shows scalar field $\phi$ versus time at the same time range.}\label{fig6}
\end{figure}

We proceed with the second case. We would like to do it for ``dark energy dominated era". For this purpose, we want to take the form of $\phi(t)$ and $A(t)$ arbitrary. However we have room for choosing the form of $a(t)$, but we want it to arise from the heart of equations spontaneously for comparing with the known scale factor for dark energy dominated era ($i.e$. $a_{D.E.} =e^{-1} e^{H_{0} t} $. Here, the coefficient $H_{0}$ in the exponential, the Hubble constant, is about $7.25 \times 10^{-11}$  $s^{-1}$, and the coefficient $e^{-1}$ is for normalizing the scale factor to $1$ at present time). We take $\phi(t)$ with decreasing nature function as
\be \label{phi- with B.T. for set2}
\phi(t) = \frac{2.708908423 \times 10^5}{\sqrt{t}},
\ee
and the form of $A(t)$ as obtained from the NS-results (Eq.(\ref{Sol. for.Eu.lag.Eq. for cyc.}))
\be \label{A- with B.T. for set2}
A(t) = - 7.068617997 \times 10^{-14} t^{1/3},
\ee
Solving Eq. (\ref{sol-FE-b.t.2}) with Eq. (\ref{phi- with B.T. for set2}) and (\ref{A- with B.T. for set2}) numerically as a boundary value problem in the dark energy dominated era's time range, [9.8 Gyr, 13.8 Gyr], with these selections
\be \label{initial for B.T. for set2}
c_{0} =1 \qquad , \qquad c_{1} = 0 \qquad , \qquad a(9.8 \times 10^9) = 0.7304711450 \qquad , \qquad a(13.8 \times 10^9) = 1,
\ee
shows deep compatible results with observational data. We present Figures (\ref{fig5}) and (\ref{fig6}) to demonstrate the results obtained. Figure (\ref{fig5}) (A) indicates the behaviors of the obtained scale factor, $a_{B.N.S.}(t)$, and the known scale factor for dark energy dominated era, $a_{D.E.}(t)$ versus time. It shows a good agreement between $a_{B.N.S.}(t)$ and $a_{D.E.}(t)$. However, maybe, it is not a well comparison, for we studied those versus time. For this reason, we study $a_{B.N.S.}(t)$ and $a_{D.E.}(t)$ versus their own redshifts. Plots overlapping in Figure \ref{fig5} (B) shows a perfect agreement between $a_{B.N.S.}(t)$ and $a_{D.E.}(t)$  versus $z_{B.N.S.}$ and $z_{D.E.}$ respectively. Figure (\ref{fig6}) (C) shows the detractive behavior of scalar field $\phi$ versus time which leads to decreasing $V(\phi)$ and incremental $f^2(\phi)$ versus $\phi$ in Figures \ref{fig6} (A) and (B) respectively. Here, $f(\phi)$ is pure imaginary again, and as mentioned above, it's not problematic.\\
As an example, one can take $a(t) = c_{1}\sinh(c_{2}t)^{2/3}$, and describe the elaborations of cosmic evolution from matter dominated era till now.

\section{WDW-Equation \label{V}}
Let us proceed with Eq.(\ref{lag. for cyc.}). So, we have the following Hamiltonian
\be \label{Hamil.}
\mathcal{H} = -\frac{3}{16} \Pi_{u} \Pi_{w} + 4 V_{0} u^2 + 2 f_{0}^{-2} u^{-2/3} \Pi_{A}^2,
\ee
where $\{\Pi_{j} = \partial \mathcal{L} / \partial \dot{Q^j} \quad ; Q^j\in \{ w, u, v = A \} \}$ are conjugated momenta of configuration
space. By a straightforward canonical quantization procedure, we have
\be \label{quan.}\begin{split}
&\Pi_{j} \rightarrow \hat{\Pi}_{j} = - i \partial_{j}, \\
&\mathcal{H} \rightarrow \hat{\mathcal{H}}(Q^j, - i \partial_{Q^j}) .
\end{split} \ee
Hamiltonian constraint gives the Wheeler-De Witt equation
\be \label{WDW}
\left[ -\frac{3}{16} (- i \partial_{u})(- i \partial_{w})+ 4 V_{0} u^2 + 2 f_{0}^{-2} u^{-2/3}(- i \partial_{A})^2
\right]|\Psi(w,u,A)\rangle =0,
\ee
in which $|\Psi(w,u,A)\rangle$ is the wave function of the universe. Pursuant to Noether symmetry, if we use the following two conserved currents
\be \label{conserved currents of SNS. used in WDW}
\Pi_{w} = \Sigma_{1} \qquad , \qquad \Pi_{A} = \Sigma_{2},
\ee
then, according to \cite{13}
\be \label{psi}
|\Psi\rangle = \sum_{j=1}^{m} e^{i \Sigma_{j}Q^j} |\chi (Q^l)\rangle , \qquad m < l \leq n ,
\ee
where $m$ is the number of symmetries, $l$ are the directions where symmetries do not exist, $n$ is the total dimension of minisuperspace, we have
\be \label{wave function}
|\Psi(w,u,A)\rangle = e^{i \Sigma_{1} w} e^{i \Sigma_{2} A} |\Theta(u)\rangle.
\ee
Note that the presence of the exponential functions is due to the separation of variables in Eq. (\ref{WDW}) and the quantum version of constraints (\ref{conserved currents of SNS. used in WDW}) which are
\begin{equation*}
-i \partial_{w} = \Sigma_{1} |\Psi(w,u,A)\rangle, \qquad -i \partial_{A} = \Sigma_{2} |\Psi(w,u,A)\rangle.
\end{equation*}
After putting this solution in Eq. (\ref{WDW}) and solving it, we get this perfect solution
\be \label{perfect wave function}
|\Psi(w,u,A)\rangle = c_{1} e^{i \Sigma_{1} w} e^{i \Sigma_{2} A} e^{\frac{b_{2} u^3 + 9 b_{3} u^{1/3}}{-3 b_{1}}}
\ee
where $b_{1}=3 i \Sigma_{0} / 16 $, $b_{2}=4 V_{0}$, $b_{3}=2 f_{0}^{-2} u^{-2/3} \Sigma_{1}^2 $, and $c_{1}$ is an integration constant. It is clear that oscillating feature of the wave function of the universe recovers the so-called Hartle criterion \cite{53}.
\section{CONCLUSION \label{VI}}
In this paper, we studied an original action in Teleparallel gravity which has never been introduced in the literature. However, the $f(R)$ version of the action (\ref{action}) has been introduced and studied in some papers \cite{45}-\cite{48}. By the use of Noether approach, we found that late--time--accelerated expansion is realized with this model. Our data analysis showed that the age of the universe is $13.86$ $Gyr$, the present amount of the scale factor, deceleration, and EoS parameters are $a_{0} = 1.00$, $q_{0} = -1$ and $W_{eff}= -1$, respectively and the scalar field, $\phi$, and coupling function, $f^2(\phi)$, with decreasing nature while scalar potential, $V(\phi)$, increasing with time. Considering deceleration parameter, we learned that the universe starts accelerating from $z = 0.524$, which is equivalent to $t_{ac.} = 6.294Gyr$. The resulting model crosses the phantom divide line from the quintessence phase to the phantom phase. By data analysis, we obtained the quadratic coupling function, $f^2(\phi)$, to be real in the action (\ref{action}) but $f(\phi)$ itself was pure imaginary. In other words, we found (by data analysis) that $f(\phi)$ has the form $f(\phi)= (\sqrt{-1}) N(\phi)= i N(\phi)= f_{0}N(\phi)$ in which $N(\phi)$ is a real function of the scalar field $\phi$ (i.e. $N(\phi) = \exp\left[\sqrt{6} \phi/(-12) \right]$ (See Eqs. (\ref{inv. inv. sol. for cyc.3.}) and (\ref{selection for constants})), but it is not problematic and is compatible with observations, since according to the action (\ref{action}), we see that $f_{0}$ appears with power two (i.e. $f^2_{0}$) in the action and also in the relevant equations (see Eqs. (\ref{FE1} - \ref{EoS}), (\ref{Eu.lag.Eq. for cyc.}), and (\ref{M2} - \ref{M42})). We showed that the obtained results satisfied Maxwell's equations in curved space-time. The amounts of electric and magnetic fields fall off with time and also they have the same norm at present time. We presented an original approach as ``B.N.S. Approach" for easy solving the field equations with keeping almost all conserved currents which we want to have, and showed that unlike Noether approach, we have solutions with this approach. Our solutions could describe dark energy dominated era. And in the section \ref{V} we considered WDW-equation and showed that wave function of the universe has oscillating features which in the cosmic evolution recovers the so-called Hartle criterion.\\

Finally, we would like to compare our NS-results with SNS-results of Ref. \cite{47} which considered $f(R)$ version of the action (\ref{action}). By taking the incorrect vector field as $A_{\mu} = (0; 0, 0, A(t))$, they showed that the scale factor is about $a(t) \thicksim (t^4 +t^{4/3} +t)^{1/3}$ while in our case it is about $a(t) \thicksim (t^4 + t^{4/3} + t + t^2)^{1/3}$. According to the extra term, $t^2$, in the parenthesis, other things are different. However, regarding the FLRW spacetime, we took different vector field as (\ref{four vector}). Anyway, both versions, show late--time--accelerated expansion. It is worth to note that the behavior of scalar field is different; that is, in our case, it has decreasing nature while in that paper it has increasing nature after a little detractive behavior. In that paper, they studied qualitative behavior of $a(t)$ and $\phi(t)$ versus time only, so we can't discuss more. Howbeit, I am sure that in that paper, one can show pure imaginary nature of $f(\phi)$ by data analysis, such as our case. However we have $f^2(\phi)$ in that action, so it is not problematic.\\

\end{document}